# Unexpected Giant Superconducting Fluctuation and Anomalous Semiconducting Normal State in $NdO_{1-x}F_xBi_{1-y}S_2$ Single Crystals

Jianzhong Liu[1†], Delong Fang[1†], Zhenyu Wang[2†], Jie Xing[1], Zengyi Du[1], Xiyu Zhu[1*], Huan Yang[1*], Hai-Hu Wen[1*]

[1] Center for Superconducting Physics and Materials, National Laboratory of Solid State Microstructures and Department of Physics, National Center of Microstructures and Quantum Manipulation, Nanjing University, Nanjing 210093, China

[2] National Laboratory for Superconductivity, Institute of Physics and National Laboratory for Condensed Matter Physics, Chinese Academy of Sciences, Beijing 100190, China

[†] These authors contributed equally to this work

Correspondence and requests for materials should be addressed to XYZ, HY, HHW (emails: zhuxiyu@nju.edu.cn, huanyang@nju.edu.cn, hhwen@nju.edu.cn).

**The $BiS_2$-based superconductors were discovered recently[1-2]. The superconductivity has been proved by many other groups[3-10]. Since the previous experiments were all done on polycrystalline samples, therefore there remains a concern whether the superconductivity is really derived from the materials intrinsically or from some secondary phases[11]. Experiments on single crystals are highly desired. In this paper, we report the successful growth of the $NdO_{1-x}F_xBi_{1-y}S_2$ single crystals. Resistive and magnetic measurements reveal that** **the bulk superconducting transition occurs at about 5 K, while an unexpected giant superconducting fluctuation appears at temperatures as high as 2-4 $k_BT_C$. Analysis based on the anisotropic Ginzbaug-Landau theory gives an anisotropy $\gamma = \sqrt{m_c / m_{ab}} \approx 30 \sim 45$. Two gap features with magnitudes of about 3.5±0.3 meV and 7.5±1 meV were observed by scanning tunneling spectroscopy. The smaller gap is associated with the bulk superconducting transition at about 5 K yielding a huge ratio $2\Delta_s^1 / k_B T_c$ =16.8, the larger gap remains up to about 26 K. The normal state recovered by applying a high magnetic field shows an anomalous semiconducting behavior. All these suggest that the superconductivity in this newly discovered superconductor cannot be formatted into the BCS theory.**

In the Bardeen-Cooper-Schrieffer (BCS) theory, the electron pairing and condensation occur simultaneously at $T_c$. The superconducting fluctuation may extend to above $T_c$ in a very narrow region (less than 10% $T_c$)[12]. Therefore it is a surprise that the superconducting fluctuation appears at temperatures far above $T_c$ in the cuprates[13-16]. In some thin but dirty metallic films, people were pursuing a man-made wide superconducting fluctuation region in terms of the Cooper pair gas state[17-18]. In addition, in the BCS scheme, the normal state after the superconductivity is suppressed should show a metallic behavior. Any semiconducting or insulating behavior appearing in the normal state should have its special reasons. Within the weak electron-phonon coupling BCS picture, one has $2\Delta_s / k_B T_c$ =3.5 with $\Delta_s$ the superconducting gap. This ratio can be slightly higher in the conventional superconductors with strong coupling, but can reach about 8 in the cuprates[19]. In this paper we will show that all three features mentioned above for the BCS scheme will be violated in the newly discovered superconductor $NdO_{1-x}F_xBi_{1-y}S_2$ single crystal samples.

The single crystals $NdO_{1-x}F_xBi_{1-y}S_2$ were grown using flux method with KCl/LiCl as the flux. The details about the sample growth are given in Methods. The crystals are very shiny with the dark-grey color. In Fig.1**b**, we show the Laue diffraction pattern on one crystal. The clear and symmetric spots can be well fitted to the model calculations of crystallography, indicating high quality of the crystals. From the data of Laue diffraction pattern, we can determine the in-plane crystalline axes, which is very helpful for the precise in-plane resistive measurements. In Fig.1**c**, we show the X-ray diffraction (XRD) patterns for the samples grown with different nominal concentrations of fluorine. It is clear that only (*00l*) reflections can be observed yielding a *c*-axis lattice constant $c$ = 13.49±0.04 Å. In Fig.1**d**, we show a picture of one crystal on which the composition is analyzed using the energy-dispersion-spectrum (EDS). From the EDS data, one can see that our crystal has a formula like $NdO_{1-x}F_xBi_{0.84}S_{1.94}$. Since the oxygen and fluorine are both light elements, the value given here about them are not reliable, although the nominal compositions of them are well documented. We found that the samples with the nominal fluorine concentration less than 30% are not or bad superconductive. Due to the error bars of the EDS measurements, we can conclude that the composition of sulfur here is close to 2, but Bi is slightly deficient. We have done the EDS analysis on about 20 crystals, the results are quite convergent and given in the Supplementary Information.

In Fig.2**a** and 2**b**, we present a typical set of resistive data measured at zero and different magnetic fields with the configurations of: (a) $H||c\text{-}axis\perp I$ and (b) $H||I||a\text{-}axis$ with the current always flowing along a-axis. The inset in Fig.1**a** gives the DC magnetization transition measured at H = 1 Oe. Interestingly the superconductivity is strongly suppressed by the magnetic field when the field is along *c*-axis, but quite robust with an in-plane field. When H||*c*-axis, the bulk superconductivity can be easily suppressed down to 2 K with only a field of 0.6 T. However, the superconductivity is very robust with H||*a*-axis and keeps presence above 2 K when the field is up to 9 T. We extract the resistive data between 20-30 K down to low temperature region as the normal state value $\rho_n(T)$, and use the different criterions of 50%, 90% and 98%$\rho_n$ for determining the upper critical field $H_{c2}(T)$. The data are shown in Fig.2**d**. One can see that the $H_{c2}(T)$ determined with above three different criterions are quite different for the cases of $H||c\perp I$ and $H||I||a$. From these data, one can determine the anisotropy by $\gamma = [dH_{c2}^{ab}(T)/dT]/[dH_{c2}^{c}(T)/dT]$ near $T_c$, we find that $\gamma$ falls into the region of 30-45. One can of course use the Werthamer-Helfand-Hohenberg (WHH) formula[20] $H_{c2} = -0.69T_c[dH_{c2}/dT]_{T_c}$ to determine the upper critical field at zero temperature. Taking the data with the criterion of 90%$\rho_n$(T), we have $T_c$ = 4.83 K, $dH_{c2}^{ab}(T)/dT = -12T/K$, $dH_{c2}^{c}(T)/dT = -0.25T/K$, finally we have $H_{c2}^{ab}(0) = 40T$ and $H_{c2}^{c}(0) = 0.833T$. This huge anisotropy has only been observed in the cuprate Bi-2212 and Bi-2223 systems. Josephson vortices are certainly expected in this new $BiS_2$-based superconducting systems.

There is an uncertainty in determining the anisotropy in above process because (1) the upper critical field is dependent on the criterion of the resistivity and (2) the curve $H_{c2}^{ab,c}(T)$ has a curvature near $T_c$ which prohibits to have a precise determination of the ratio $\gamma = [dH_{c2}^{ab}(T)/dT]/[dH_{c2}^{c}(T)/dT]$. There is actually an alternative way to determine the anisotropy, especially for one fixed temperature. According to the anisotropic

Ginzburg-Landau theory, the resistivity in the mixed state depends on the effective filed $H/H_{c2}^{GL}(\theta)$ with θ the angle enclosed between *c*-axis and the field direction. The effective upper critical field $H_{c2}^{GL}(\theta)$ at an angle θ is given by

$$H_{c2}^{GL}(\theta) = H_{c2}^{c} / \sqrt{\cos^2(\theta) + \gamma^{-2}\sin^2(\theta)} \ . \qquad (1)$$

Therefore by using a scaling variable $H/H_{c2}^{GL}(\theta) = H\sqrt{\cos^2(\theta) + \gamma^{-2}\sin^2(\theta)}$ , the resistivity measured at different magnetic fields but at a fixed temperature should be scalable to one curve[21]. In Fig.3**a**, **b** and **c**, we show the angle dependence of the in-plane resistivity at 3, 3.5 and 4 K. Now the measuring current is flowing along *a*-axis, the magnetic field is always perpendicular to the current direction. Fig.3**d** presents the scaling behavior according to above mentioned method. One can see that the scaling quality looks quite good for all three temperatures 3, 3.5 and 4 K, yielding the anisotropy of 36, 31 and 30 respectively. Therefore we can safely conclude that the anisotropy of this new superconducting system is about 30-45.

Another interesting discovery on the single crystals is the giant superconducting fluctuation effect. In Fig.2**a**, if we extrapolate the resistivity curve between 20-30 K linearly down to low temperatures, one can see an enhanced excess conductivity clearly above $T_c$. To illustrate this, in inset of Fig.2**b**, we present the difference of the resistivity between the measured data ($\rho=1/\sigma$) and the extrapolated linear line ($\rho_n=1/\sigma_n$) $\Delta\rho=\rho_n-\rho=1/\sigma_n-1/\sigma=\Delta\sigma/\sigma\sigma_n$, with σ, $\sigma_n$ and Δσ the measured, extrapolated and the excess conductivity, respectively. It is clear that the resistivity difference vanishes at about 15 K. In Fig.2c, we show the resistivity curve on another sample, it is clear that the superconducting fluctuating region extends up to about 20 K, as marked by $T*$.

In order to check whether we really have a strong superconducting fluctuation, we measured the in-plane resistivity with a rotating in-plane magnetic field. The measuring current is along a- (or b-) axis, and the sample is rotated with an in-plane magnetic field. The angle enclosed between the current direction and the magnetic field is $\phi$. When $\phi=0=\pi$, the Lorenz force on the vortices is zero and thus the dissipation is the minimum, while it becomes maximum with $\phi=\pi/2=3\pi/2$. Interestingly the same two fold oscillations appear from 4 K all the way up to 10 K. At 4K, we have strong reason to believe that the vortex motion dominates the dissipation. From the systematic evolution from 4 K to 10 K, we think that the 2-fold resistivity oscillation is still induced by the vortex motion, even at 10 K. This leads to the conclusion that the superconducting fluctuation exists at least up to 10 K from the angle dependent resistive data, although the bulk transition occurs at $T_c$ = 5K.

In order to further unravel the superconductivity in the sample, we have measured the scanning tunneling spectroscopy (STS) on the single crystals. The measurements were done with a commercial instrument Unisoku-11T with ultra-high vacuum system. The single crystals were cleaved under an ultra-high vacuum then followed by the measurements at a low temperature down to 0.4 K. The details of the measurements and results are given in Methods. One typical set of data measured at temperatures from 0.4 to 26 K is shown in Fig.5**a**. One can see that, the data measured at low temperatures exhibit clearly two sets of energy gaps, one small gap $\Delta_s^1 \approx 3.5 \pm 0.3$ meV and a larger gap $\Delta_s^2 \approx 7.5 \pm 1$ meV. The small gap is marked by the dashed red line in Fig.5**b**. Interestingly, at the bias voltage of the small gap, two coherence peaks appear at about 3.5 meV and gradually disappear at about 6 K. Therefore we can conclude that this small gap seems corresponding to the bulk superconducting transition at around 5 K. This feature can be easily observed in the STS with the data normalized by that at 8 K and shown in Fig.5**c**. As shown in Fig.5**b**, the larger gap, however, shows up as the energy features at about 7.5±1 meV at low temperatures, and as a suppression to the deferential conductance (*dI/dV*) or the quasiparticle density of states (DOS) at temperatures above 8 K. This feature can extend to high temperatures and gradually vanishes at about 26 K. One can naturally connect this larger gap to the giant superconducting fluctuations in the resistive measurements.

In the resistive data shown in Fig.2**a**, one can see that the normal state reveals a semiconducting behavior when superconductivity is completely suppressed. Similar phenomenon was observed and emphasized in the polycrystalline $CeO_{1-x}F_xBiS_2$ samples previously[5,6]. This is anti-intuitive to the BCS picture, since a metallic state should be recovered in the normal state. One may argue that this semiconducting behavior is induced by the localization effect of electrons in the low dimensional system as derived from the band structure calculations[22,23]. This scheme is however also not workable here. For a low dimensional system, impurities may induce a semiconducting behavior due to the localization effect, the magnetic field will weaken this localization effect and the semiconducting feature. While our observation is that, when the superconductivity is suppressed, the semiconducting feature above $T_c$ shows up. In addition, the magnetic field seems enhancing this semiconducting effect in a certain field region. Another possibility is the possible Kondo-like scattering which will induce a semiconducting behavior. This explanation can also be excluded because we don't have this type of strong magnetic impurities in the samples. In addition, the weakening effect of the semiconducting behavior by the magnetic field in a Kondo system is not observed here. For this low dimensional system, there is however a tendency for the charge-density-wave (CDW). When the superconductivity is suppressed, the CDW tendency will be promoted and lead to a semiconducting feature. While the interesting point is that, the "normal state" here above $T_c \approx 4.83$ K at zero field seems not showing even a weak or negligible semiconducting effect. If it were induced by the CDW tendency, it would show out above $T_c$ even the field is zero. Therefore this semiconducting feature of the normal state under a magnetic field is non-trivial and requires further input.

From the STS measurements we observed two gap features at different energies. The smaller gap $\Delta_s^1 \approx 3.5 \pm 0.2$ meV is corresponding to the bulk superconducting transition at about 5 K. Taking $T_c$ = 4.83 K, the ratio $2\Delta_s^1 / k_B T_c$ = 16.8! This is much larger than the theoretical value 3.5 in the weak coupling limit of the BCS model. If we use the upper boundary of the superconducting fluctuation temperature, for example 20 K as the pairing temperature, we have $2\Delta_s^1 / k_B T_c$ =4.06, which is not far from the BCS value 3.5. This probably suggests that the superconducting pairing occurs at a high temperature (for example 20 K). Because of the very low dimensionality, the superconducting fluctuation is very strong, and the bulk transition at about 5 K is actually corresponding to the phase coherence temperature. This gives strong resemblance to the case in cuprate superconductors. Following this line, the gap feature at $\Delta_s^2 \approx 7.5 \pm 1$ meV may be a pseudogap, which may be induced by the tendency of the CDW order, or by the local pairing governed by the valence fluctuation effect of the $Bi^{2+}$ and $Bi^{3+}$ ionic states[24]. If the latter mechanism is proved, this interesting $BiS_2$-based superconductor will provide us a new platform for the novel pairing mechanism of the valence fluctuation. The recent theoretical calculations suggest a significant band splitting due to the spin-orbital coupling effect in the $BiS_2$ based superconductors[23]. In this case, one can expect a mixture of the spin singlet and spin triplet[25] in the system. The clear violation of the predictions of the BCS picture for conventional superconductors observed in this work, i.e., the giant superconducting fluctuation, the huge gap feature and the unexpected semiconducting behavior after the superconductivity is suppressed, will certainly stimulate the enthusiasm of research in this newly discovered superconducting system, especially on single crystal samples[26].

## Methods

### I. Sample growth

The single crystals were grown by flux method using KCl/LiCl (molar ratio KCl : LiCl = 3 : 2) as the flux. Firstly, the polycrystalline samples were grown by a solid-state reaction. In growing the polycrystalline samples, the starting materials Nd grains (purity 99.9%), $NdF_3$ (purity 99.9%), $Nd_2O_3$ (purity 99.9%), $Bi_2S_3$ (purity 99.9%) and S (purity 99.9%) were mixed in stoichiometry as the formula $NdO_xF_{1-x}BiS_2$. Then we pressed the mixture into a pellet shape and sealed it in an evacuated quartz tube. It was heated up to 750°C for 10 hour. The resultant pellet was smashed and grounded together with KCl/LiCl powder of molar ratio (KCl/LiCl : $NdO_{1-x}F_xBiS_2$ = 25 : 1) and sealed in an evacuated quartz tube. The mixed powder was heated up to 750°C for 5 days followed by cooling down at a rate of 5°C /hour to 450°C.

After the samples was cooled to room temperature by shutting off the power of the furnace, we used water to dissolve the flux and got single crystals with lateral sizes of about 1 mm and thickness of about 10-100 μm.

**II. STM measurements**

The scanning tunneling spectra (STS) were measured with an ultra-high vacuum, low-temperature, and high-magnetic-field scanning probe microscope USM-1300 (Unisoku Co., Ltd.). The samples were cleaved at room-temperature in ultra-high vacuum with a base pressure about $1\times10^{-9}$ torr. The cleaved surfaces were always flat and shiny, and the atomically resolved image can be obtained sometimes. However the surfaces were unstable and the spectra showed a semiconductor behavior. This may be originated from the polar BiS top surface after the cleavage. The cleaved surfaces were then cleaned by argon iron bombardment with an acceleration voltage of 1500 V to get a more stable surface. We can obtain the superconducting spectra on such surface after the treatment. In all STM/STS measurements, Pt/Ir tips were used. To lower down the noise of the differential conductance spectra, a lock-in technique with an ac modulation of 1 mV at 987.5 Hz was used.

## References and Notes

1. Mizuguchi,Y.,Fujihisa,H., Gotoh,Y., Suzuki,K.,Usui,H., Kuroki,K., Demura,S., Takano,Y., Izawa,H & Miura, O. $BiS_2$-based layered superconductor $Bi_4O_4S_3$. *Phys. Rev. B* **86**, 220510 (2012).
2. Mizuguchi, Y., Demura, S., Deguchi, K., Takano, Y, FujihisaH., Gotoh, Y., Izawa, H &

Miura, O. Superconductivity in Novel $BiS_2$-Based Layered Superconductor $LaO_{1-x}FxBiS_2$. *J. Phys. Soc. Japan* **81**, 114725 (2012).

3. Li, S., Yang, H., Fang, D. L., W, Z. Y., Tao, J., Ding, X. X.& Wen, H. H. Strong coupling superconductivity and prominent superconducting fluctuations in the new superconductor $Bi_4O_4S_3$. *Sci China-Phys Mech Astron* **56**, 2019-2025(2013).
4. Singh, S. K., Kumar, A., Gahtori, B., Kirtan, S., Sharma, G., Patnaik, S & Awana, V. P. S. Bulk Superconductivity in Bismuth Oxysulfide $Bi_4O_4S_3$. *J. Am. Chem. Soc.***134**, 16504-7 (2012).
5. Xing, J., Li, S., Ding, X. X., Yang, H. & Wen, H. H. Superconductivity appears in the vicinity of semiconducting-like behavior in $CeO_{1-x}F_xBiS_2$. *Phys. Rev. B* **86,** 214518 (2012).
6. Kotegawa, H., Tomita, Y., Tou, H., Izawa, H., Mizuguchi, Y., Miura, O., Demura, S., Deguchi, K. & TakanoY. Pressure Study of $BiS_2$-Based Superconductors $Bi_4O_4S_3$ and La(O,F)$BiS_2$. *J. Phys. Soc. Japan* **81**, 103702(2012).
7. Tan, S. G., Tong, P., Liu, Y., Lu, W. J., Li, L. J., Zhao, B. C.& Sun, Y. P. Suppression of superconductivity in layered $Bi_4O_4S_3$ by Ag doping. *Eur. Phys. J. B.* **85**, 414(2012).
8. Yazici, D., Huang, K., White, B. D., Chang, A. H., Friedman, A. J. & Maple, M. B. Superconductivity of F-substituted $LnOBiS_2$ (Ln=La, Ce, Pr, Nd, Yb) compounds. *Philos. Mag.***93,** 673-80(2012).
9. Shruti, Srivastava, P. & Patnaik, S. Evidence for fully gapped strong coupling s-wave superconductivity in $Bi_4O_4S_3$. *J. Phys.: Condens. Matter***25,** 312202 (5pp)(2013).
10. Wolowiec, C. T., Yazici, D., White, B. D., Huang, K. & Maple, M. B. Pressure-induced

enhancement of superconductivity and suppression of semiconducting behavior in $LnO_{0.5}F_{0.5}BiS_2$ (Ln = La,Ce) compounds. *Phys. Rev. B* **88,** 064503(2013).

11. Phelan, W. A., Wallace, D. C., Arpino, K. E., Neilson, J. R., Livi, K. J., Seabourne C. R., Scott, A. J. & McQueen, T. M. Stacking Variants and Superconductivity in the Bi−O−S System. *J. Am. Chem. Soc.* **135**, 5372−5374(2013).
12. Aslamasov, L. & Larkin, L. Effect of fluctuations on the properties of a superconductor above the critical temperature. *Sov. Phys. Solid State* **10**, 875 (1968).
13. Xu, Z. A., Ong, N. P., Wang Y, Kakeshita, T. & Uchida, S. Vortex-like excitations and the onset of superconducting phase fluctuation in underdoped $La_{2-x}Sr_xCuO_4$. *Nature* **406**, 486-488 (2000).
14. Ri, H. C., Gross, R., Gollnik, F., Beck, A., Huebener, R. P., Wagner, P. & Adrian, H. Nernst, Seebeck, and Hall effect in the mixed state of $YBa_2Cu_3O_{7-\delta}$ and $Bi_2Sr_2CaCu_2O_{8+\delta}$: a comparative study. *Phys. Rev. B* **50**, 3312-3329 (1994).
15. Deutscher, G. Coherence and single-particle excitations in the high-temperature superconductors. *Nature* **397**, 410-412(1999).
16. Tesanovic, Z. Extreme type-II superconductors in a magnetic field: A theory of critical fluctuations. *Phys. Rev. B* **59**, 6449-6474 (1999).
17. Haviland, D. B., Liu, Y. & Goldman, A. M. Onset of superconductivity in the two-dimensional limit. *Phys. Rev. Lett.* **62**, 2180 (1989). Liu, Y., Haviland, D., Nease, B & Goldman, A. Insulator-to-superconductor transition in ultrathin films. *Phys. Rev. B* **47**, 5931(1993).
18. Mason, N. & Kapitulnik, A. Dissipation Effects on the Superconductor-Insulator

Transition in 2D Superconductors. *Phys. Rev. Lett.* **82**, 5341–5344 (1999)

19. Hudson, E. W., Pan, S. H., Gupta, A. K., Ng, K. W., Davis, J. C. Atomic-Scale Quasi-Particle Scattering Resonances in $Bi_2Sr_2CaCu_2O_{8+\delta}$. *Science* **285**, 88-91(1999). See also Ali Yazdani, A., Howald, C. M., Lutz, C. P. Kapitulnik, A. & Eigler, D. M. Impurity-Induced Bound Excitations on the Surface of $Bi_2Sr_2CaCu_2O_8$. *Phys. Rev. Lett.* **83**, 176-179(1999).
20. Werthamer, N. R., Helfand, E. & Hohenberg, P. C. Temperature and Purity Dependence of the Superconducting Critical Field $H_{c2}$. III. Electron Spin and Spin-Orbit Effects. *Phys. Rev.* **147**, 295 (1996).
21. Blatter, G., Geshkenbein, V. B. & Larkin, A. I. From isotropic to anisotropic superconductors: A scaling approach. *Phys. Rev. Lett.* **68**, 875-878 (1992).
22. Usui, H., Suzuki, K. & Kuroki, K. Minimal electronic models for superconducting $BiS_2$ layers. *Phys. Rev. B* **86**, 220501(R)(2012).
23. Wan, X. G., Ding, H. C., Savrasov, S. Y. & Guan, C. G. Electron-phonon superconductivity near charge-density-wave instability in $LaO_{0.5}F_{0.5}BiS_2$ Density-functional calculations. *Phys. Rev. B***87**, 115124(2013).
24. Onishi, Y. & Miyake, K. *J. Phys. Soc. Jpn.* 69, 3955 (2000). Watanabe, S. Imada, M. & Miyake, K. Superconductivity emerging near quantum critical point of valence transition. *J. Phys. Soc. Jpn.* **75**, 043710 (2006).
25. Yang, Y., Wang, W. S., Xiang, Y. Y., Li, Z. Z. & Wang, Q. H. Pairing with dominant triplet component and possible weak topological superconductivity in $BiS_2$-based superconductors. Preprint at http://arxiv.org/abs/1307.2394(2013).

26. Note: During the preparation of present manuscript, we became aware of a newly posted work reporting the growth and the resistive measurements of the F-substitute $NdOBiS_2$ single crystal, see Nagao, M. *et al.* arxiv.org/abs/1309.6400(2013). The anisotropy determined from this work is about 30 which is close to ours.

**Acknowledgements** We thank Qianghua Wang, Jianxin Li and K. Kuroki for helpful discussions. This work was supported by NSF of China, the Ministry of Science and Technology of China (973 projects: 2011CBA00102, 2012CB821403, 2010CB923002) and PAPD.

**Author Contributions** The sample were grown by J.Z.L. and X.Y.Z. The transport measurements were done by J.Z.L., J.X. and H-H.W. The low-temperature STS measurements were performed by D.F., Z.W., Z.D., H.Y. & H-H.W. H-H.W. coordinated the whole work and wrote the manuscript, which was supplemented by other co-authors. All authors have discussed the results and the interpretation.

**Author Information** The authors declare no competing financial interests.

## Figures and Captions

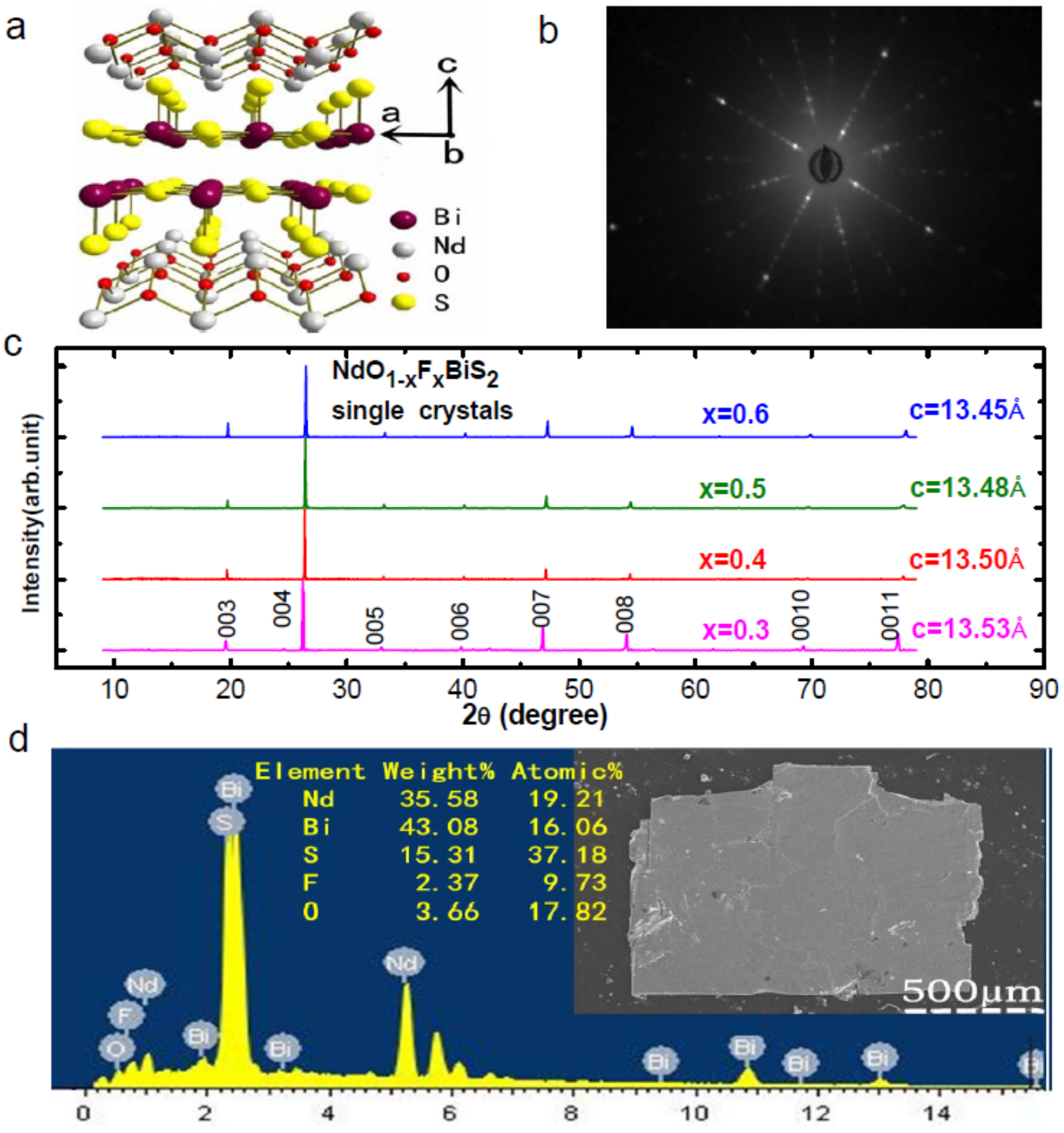

**Figure 1 | Structure characterizations of $NdO_{1-x}F_xBiS_2$ single crystals. (a,b)** The schematic show of the structure of $NdO_{1-x}F_xBiS_2$ and the back Laue X-ray Diffraction Pattern of $NdO_{0.5}F_{0.5}BiS_2$ single crystal. **(c)** X-ray diffraction pattern of $NdO_{1-x}F_xBiS_2$ single crystals. The XRD patterns exhibit the *c*-axis preferred orientation of the single crystals. Lattice parameter *c* shrinks slightly with doping more fluorine into $NdOBiS_2$. **(d)** Energy Dispersive X-ray microanalysis spectrum taken on one $NdO_{0.5}F_{0.5}BiS_2$ single crystal. Considering the oxygen inaccuracies by the method, the formula of the sample could be written as $NdO_{1-x}F_xBi_{0.84}S_{1.94}$. The inset shows the SEM photograph of the crystal with typical dimensions of $1.2\times0.8\times0.02$ $mm^3$.

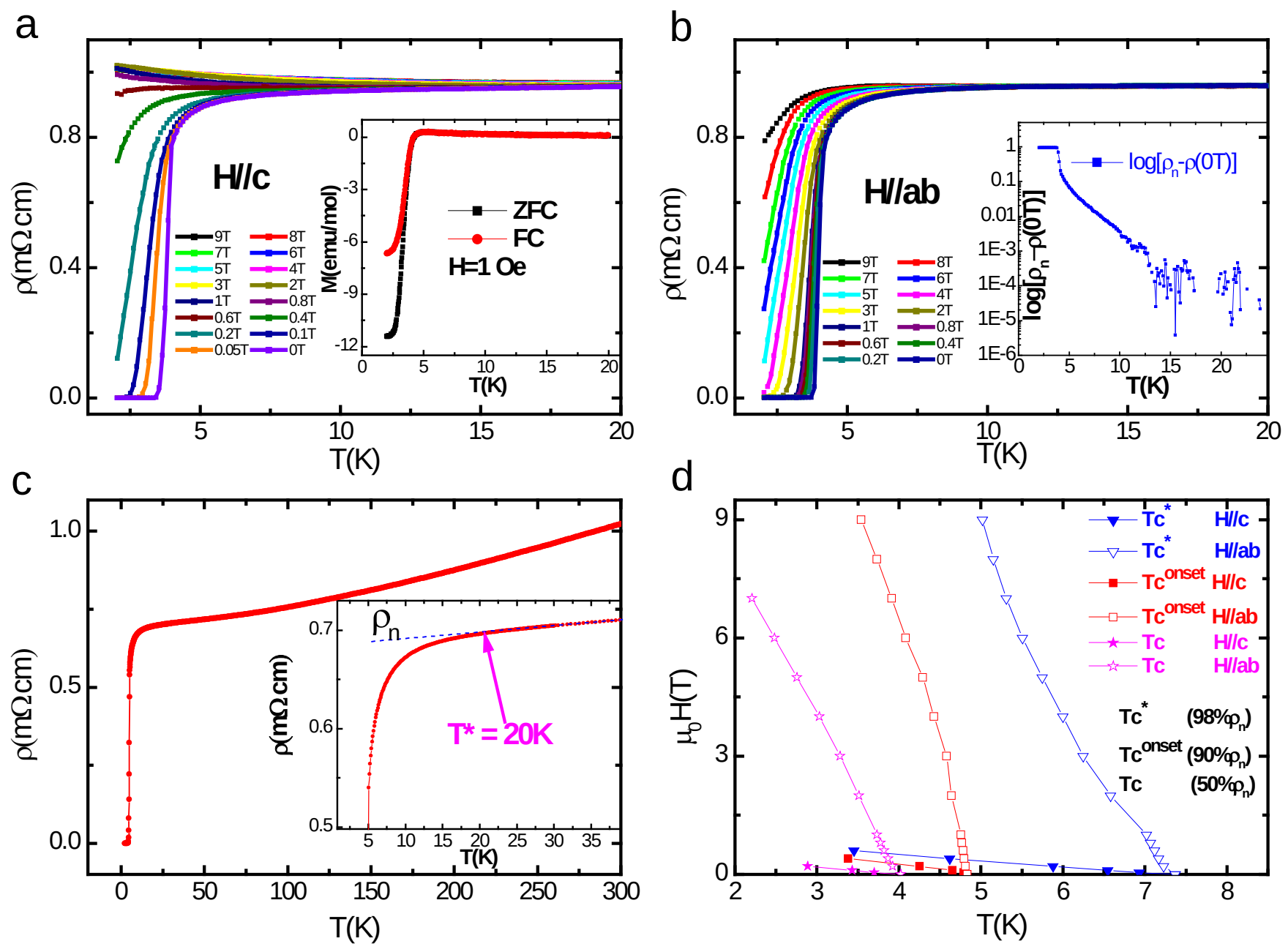

**Figure 2 | Resistivity, susceptibility and upper critical field of $NdO_{0.5}F_{0.5}BiS_2$.** (**a, b**) The temperature dependence of resistivity for the $NdO_{0.5}F_{0.5}BiS_2$ single crystal at zerofield and under magneticfields of H//c (a) and H//ab (b) up to 9 T. The inset of (a) show the Zero-field-cooled (ZFC) susceptibility and field-cooled (FC) susceptibility data at 1 Oe. The inset of (b) shows the temperature dependence of logρ= log[$\rho_n$- ρ(0T)]. (**c**) The temperature dependence of resistivity for the $NdO_{0.5}F_{0.5}BiS_2$ single crystal at zero field from 2 K to 300 K, the inset gives the superconducting fluctuation up to 20K. (**d**) Upper critical field determined using the criterion of 98%$\rho_n$, 90%$\rho_n$ and 50%$\rho_n$ of H//c and H//ab.

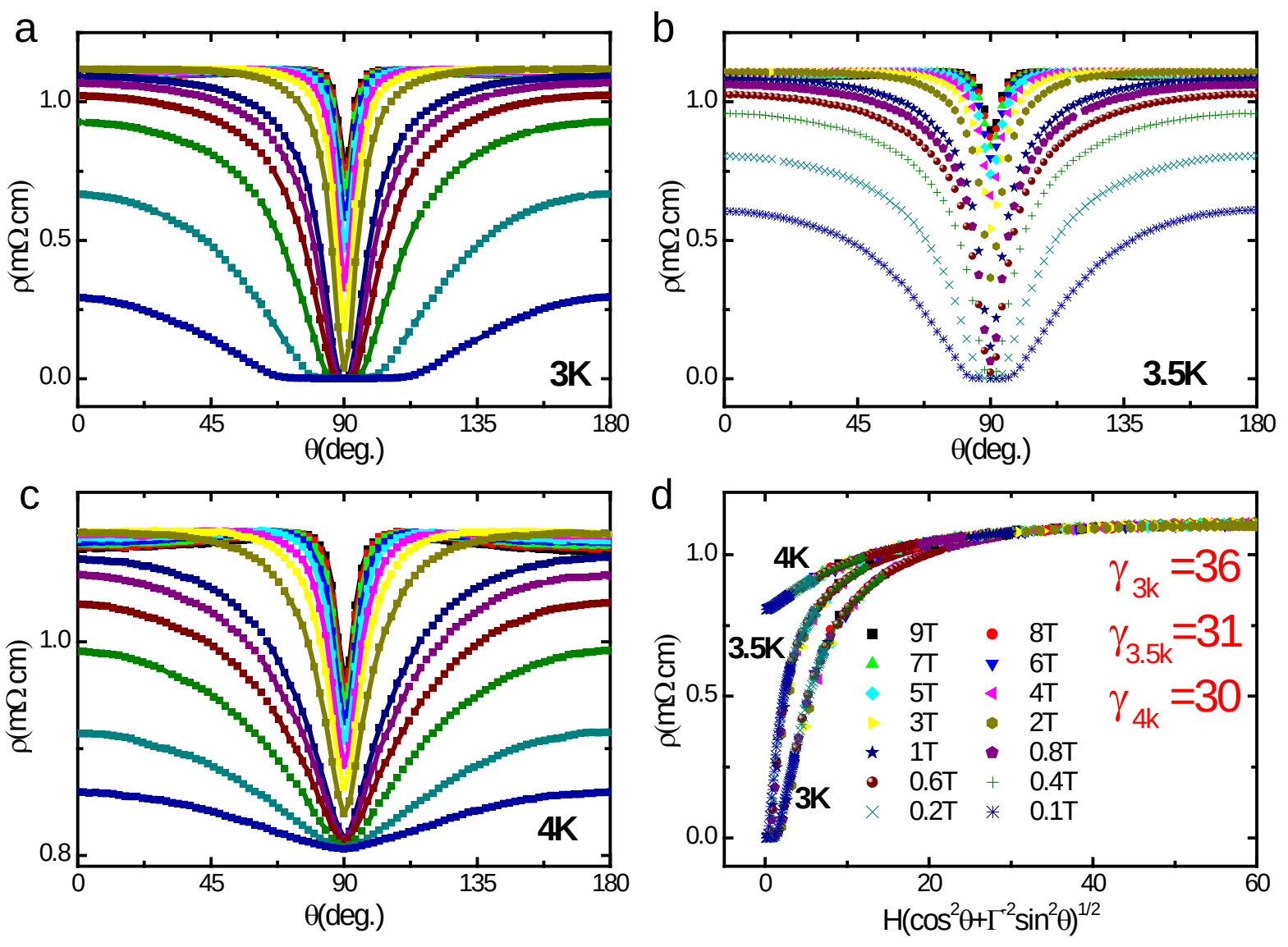

**Figure 3 | Angular dependence of resistivity and anisotropy for the $NdO_{0.5}F_{0.5}BiS_2$ single crystal.** (**a, b, c**) Angular dependence of resistivity at (a) 3 K, (b) 3.5 K and (c)4 K with $\mu_0 H$ from 0.1 T to 9 T. Here, the angular (θ) is the angle between the magnetic field and *c*-axis, with current flowing always in-plane and perpendicular to the field. (d) Scaling of the resistivity versus $\tilde{H} = H\sqrt{\cos^2(\theta) + \gamma^{-2}\sin^2(\theta)}$ at 3, 3.5 and 4 K in different magnetic fields. The data measured at a fixed temperature and different magnetic fields are scaled nicely by adjusting γ.

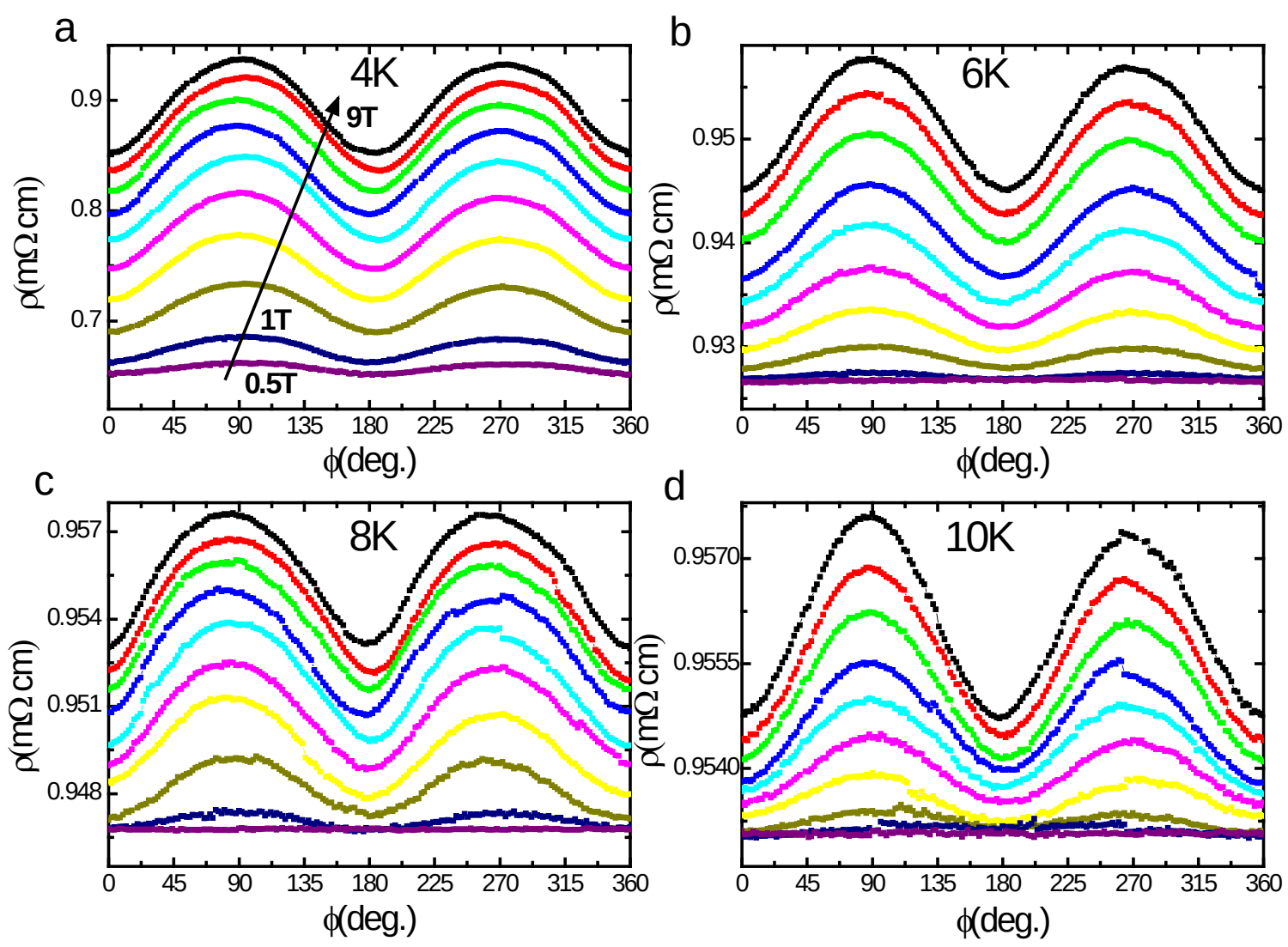

**Figure 4 | In-plane angular dependence of resistivity with H//ab and I//a-axis.** (**a, b, c, d**) Angular dependence of resistivity at (a) 4K, (b) 6 K, (c) 8K and (d)10K in $\mu_0 H$ from 0.5 T to 9 T. Here, the angle (ϕ) means the angle between the in-plane magnetic field and the direction of applied current along *a*-axis.

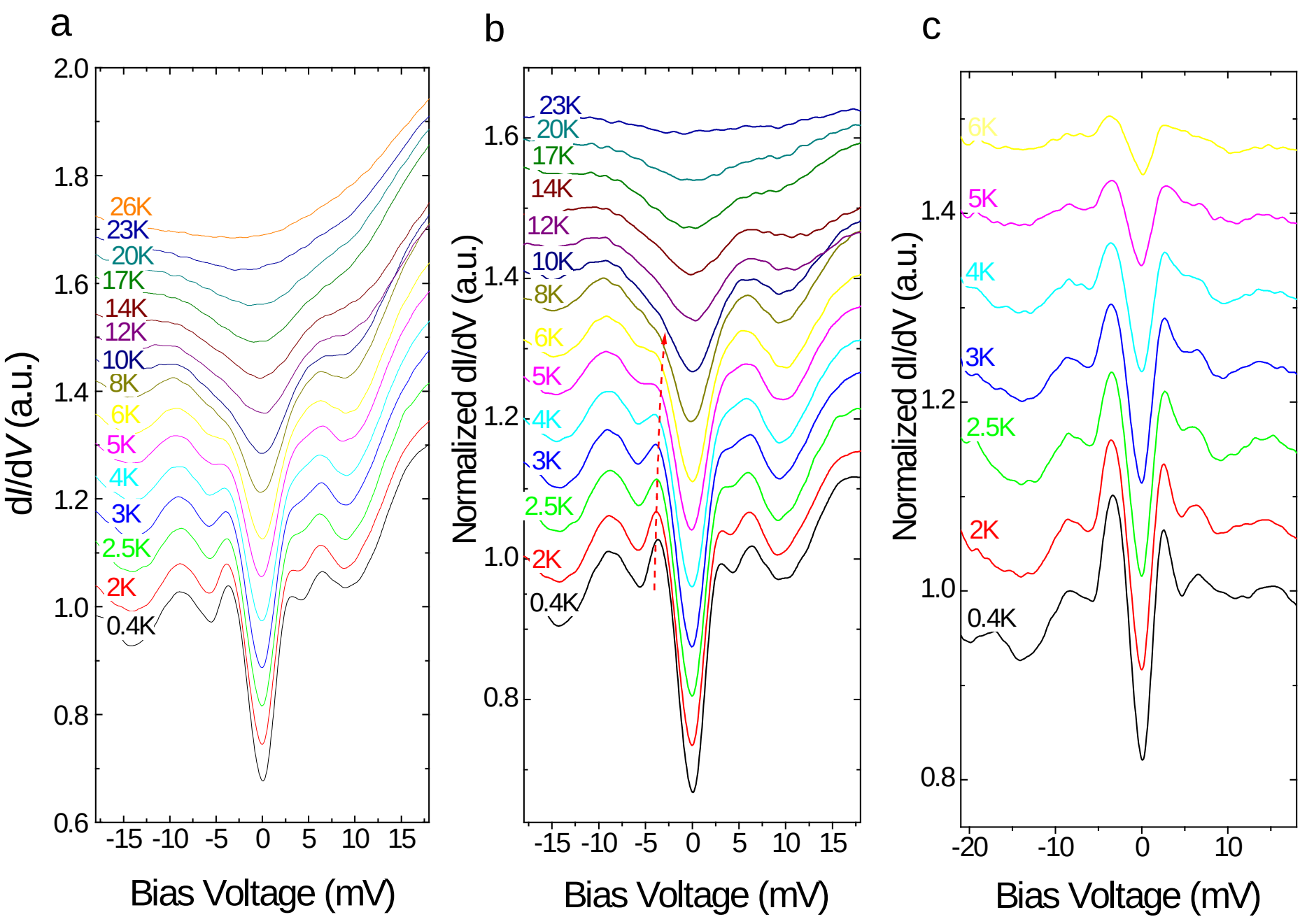

**Figure 5 | STS spectra for the $NdO_{0.5}F_{0.5}BiS_2$ single crystal. a**, The evolution of the STS spectra with temperature increased from 0.4 K to 26 K. In the low temperature region, we can see the superconducting coherence peaks emerging at about ±3.5(±0.3) meV and two higher energy hump features arising at about ±7.5(±1) meV. The larger gap features are slightly asymmetric in positions, which may be induced by the tilted background. The superconducting coherence peaks vanishes at about 6 K, while the high energy hump features are suppressed with increasing temperature and gradually vanish above 26 K. **b**, The STS spectra normalized by the one measured in the normal state (at 26 K). A dashed red color line highlights the superconducting coherence peaks at around 3.5 meV. **c**, The STS spectra normalized by the one measured in the normal state (at 8 K).

# Supplementary Information

## I. The Laue x-ray diffraction pattern

The structure and quality of the crystals were checked with the Laue x-ray diffraction. The experimental result is presented in Fig.S1**a**. One can see very symmetric and bright spots, indicating a good crystallinity. In Fig.S1b we show the model fitting results using the PSL Viewer software. The pink, blue and yellow lines represent [*01m*], [*11m*] and [*21m*] directions respectively. The horizontal and vertical directions of the pattern correspond to *a*-axis or *b*-axis.

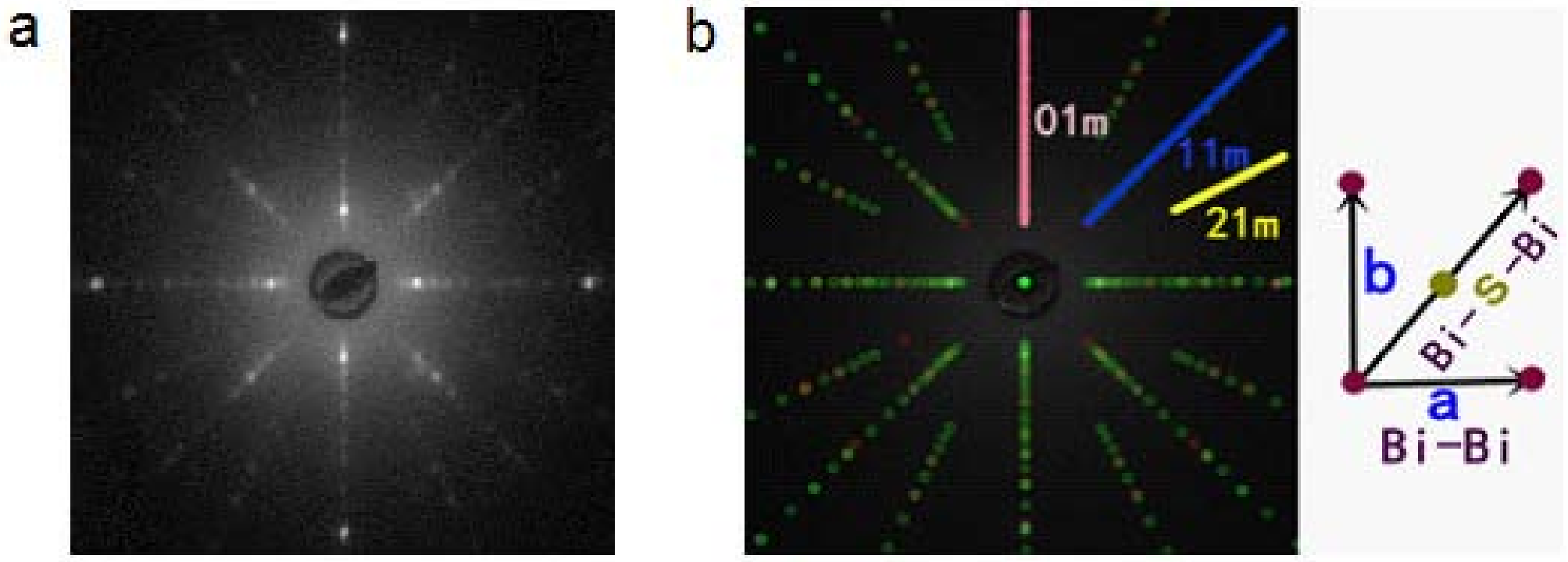

**Figure S1 | Laue diffraction pattern and analysis.** (**a**) The Laue x-ray diffraction pattern of the $NdO_{0.5}F_{0.5}BiS_2$ single crystal with dimensions of about 1000×700×15 $\mu m^3$. The inject direction of X-ray was perpendicular to the *ab* plane of the sample. (**b**) The fitting result of Laue x-ray diffraction pattern using PSL Viewer software.

## II. Composition analysis using the EDS technology

Three single crystals were randomly selected from single crystals grown with the nominal compositions of $NdO_{0.6}F_{0.4}BiS_2$ and $NdO_{0.5}F_{0.5}BiS_2$, respectively. Each sample was measured on three different positions (a, b, c) by EDS. The obtained actual compositions were normalized to Nd = 1. The last column is the averaged value of three positions for the same sample. The accurate content of O (y) couldn't be acquired by EDS, partially because of the contamination of oxygen in the air. Clearly there are some Bi and S vacancies in our single crystals. Interestingly the obtained concentration of fluorine is quite close to the nominal one.

**Table S1 | Compositions of single crystals obtained from the energy dispersive x-ray spectroscope (EDS)**

| Nominal composition | Sample | different position | Actual composition | Average composition |
|---|---|---|---|---|
| $NdO_{0.6}F_{0.4}BiS_2$ | Sample1 | a | $NdO_yF_{0.42}Bi_{0.82}S_{1.90}$ | $NdO_yF_{0.41}Bi_{0.83}S_{1.88}$ |
| | | b | $NdO_yF_{0.41}Bi_{0.83}S_{1.87}$ | |
| | | c | $NdO_yF_{0.41}Bi_{0.85}S_{1.88}$ | |
| | Sample2 | a | $NdO_yF_{0.40}Bi_{0.83}S_{1.85}$ | $NdO_yF_{0.40}Bi_{0.83}S_{1.86}$ |
| | | b | $NdO_yF_{0.42}Bi_{0.83}S_{1.86}$ | |
| | | c | $NdO_yF_{0.40}Bi_{0.83}S_{1.88}$ | |
| | Sample3 | a | $NdO_yF_{0.43}Bi_{0.85}S_{1.88}$ | $NdO_yF_{0.43}Bi_{0.85}S_{1.87}$ |
| | | b | $NdO_yF_{0.42}Bi_{0.86}S_{1.87}$ | |
| | | c | $NdO_yF_{0.45}Bi_{0.85}S_{1.86}$ | |
| $NdO_{0.5}F_{0.5}BiS_2$ | Sample4 | a | $NdO_yF_{0.46}Bi_{0.88}S_{1.91}$ | $NdO_yF_{0.48}Bi_{0.85}S_{1.93}$ |
| | | b | $NdO_yF_{0.50}Bi_{0.84}S_{1.94}$ | |
| | | c | $NdO_yF_{0.48}Bi_{0.83}S_{1.93}$ | |
| | Sample5 | a | $NdO_yF_{0.55}Bi_{0.87}S_{1.91}$ | $NdO_yF_{0.51}Bi_{0.86}S_{1.92}$ |
| | | b | $NdO_yF_{0.48}Bi_{0.86}S_{1.94}$ | |
| | | c | $NdO_yF_{0.51}Bi_{0.86}S_{1.92}$ | |
| | Sample6 | a | $NdO_yF_{0.53}Bi_{0.88}S_{1.96}$ | $NdO_yF_{0.52}Bi_{0.86}S_{1.94}$ |
| | | b | $NdO_yF_{0.55}Bi_{0.85}S_{1.94}$ | |
| | | c | $NdO_yF_{0.48}Bi_{0.87}S_{1.91}$ | |